\documentclass[letterpaper, 10 pt, conference]{ieeeconf}  

\IEEEoverridecommandlockouts                              

\usepackage{color} 

\title{\LARGE \bf
	A Neuro-inspired Theory of Joint Human-Swarm Interaction
}

\author{Jonas D. Hasbach$^{1,2}$ and Maren Bennewitz$^{2}$
	\thanks{\textcolor{red}{\textbf{This short paper was accepted and presented at the Workshop on Human-Swarm Interaction at IEEE ICRA 2020.}}}
	\thanks{$^{1}$Jonas D. Hasbach is with the Department of Human-Machine-Systems, Fraunhofer FKIE,
		Wachtberg, Germany.
		{\tt\small jonas.hasbach@fkie.fraunhofer.de}}%
	\thanks{$^{2}$Jonas D. Hasbach and Maren Bennewitz are with the Humanoid Robots Lab, Computer Science,
		University of Bonn, Germany.
		{\tt\small maren@cs.uni-bonn.de}}%
}

\begin{document}

	\maketitle
	\thispagestyle{empty}
	\pagestyle{empty}

	\begin{abstract}
		Human-swarm interaction (HSI) is an active research challenge in the realms of swarm robotics and human-factors engineering. Here we apply a cognitive systems engineering perspective and introduce a neuro-inspired joint-systems theory of HSI. The mindset defines predictions for adaptive, robust and scalable HSI dynamics and therefore has the potential to inform human-swarm loop design.
	\end{abstract}

\section{Motivation}

For the real world application of swarm robotics, human operators are required to be part of the system loop. Reasons are (1) the swarm's inability to achieve mission goals independently \cite{Conant1970a}, (2) human out of loop phenomena \cite{Bainbridge1983} as well as (3) legal and ethical concerns \cite{Verbruggen2019}. 

The objective of human-swarm interaction (HSI) \cite{Kolling2016} is to combine the distributed nature of robot swarms with the centralized control and feedback demands of humans into one loop \cite{Barca2013}. For example, swarm robots may interact locally with the operator in a fire fighting scenario \cite{Penders2011}.


HSI requires holistic theories about human-swarm loops that can be used to inform design \cite{Woods2006a}. Therefore, here we formulate a holistic neuro-inspired theory of how to best combine human- and swarm properties into one joint human-swarm loop. Testable predictions are deducted which will allow for the adjustment of the theory by empirical probing.

\section{Joint Human-Swarm Loop} 

Rather than focusing on the interaction between operator and machine, cognitive systems engineering applies a cybernetic perspective which focuses on how the operator and machine can jointly accomplish system goals \cite{Hollnagel2005}.

The system goals, and therefore the desired human-swarm-loop implementation, depend on the mission scenario. However, there are three system properties of HSI that seem desirable independent of mission goals; adaptation, robustness and scalability. We selected these variables guided by system theorist Beer's rule: \textit{'The purpose of a system is what it does'} \cite[p. 99]{Beer1985}. While adaptation is important for human-machine loops in general \cite{Woods2006a,Woods2015}, robustness and scalability have been defined as desired swarm properties \cite{Kolling2016,Barca2013}. 


\textbf{Adaptivity} is a system property that holds critical system variables in an acceptable range over different situations \cite{Ashby1960}, i.e. adaptive systems cope with surprises \cite{Woods2006a, Woods2015}. Compared to swarms, human operators are capable of much greater flexible dynamics with which they can adapt to uncertainties during missions. In contrast, \textbf{robustness} is the ability of a system to cope with demands that are expected \cite{Woods2015} (e.g. continue the mission after some swarm robots are lost). A system is \textbf{scalable} if it is capable of adjusting the number of network nodes during deployment, such as two separated bird flocks forming interactions and becoming one unity. 


Now the \textbf{challenge of HSI} (CoHSI) can be defined as \textit{joining the centralized nature of the operator with the distributed nature of the swarm \cite{Barca2013} into one goal-directed system while promoting the emergence of adaptive, robust and scalable behaviour}.

\section{Neuro-inspiration}

As humans are biological systems and robotic swarms are often bio-inspired, it seems appealing that HSI could benefit from bio-inspiration as well. Similarities between neural- and swarm principles have been discussed under the label 'swarm cognition' \cite{Trianni2011a}. In the following, we approach the CoHSI from a neuro-inspired angle.


\subsection{Swarm-Amplified Humans} In the biological nervous system, stereotypical locomotion is generated by low-level ('front end') \textbf{central pattern generators} (CPG) \cite{EveMarder2001, Ijspeert2008}. While higher-order ('back end') cognition signals may modulate the activity of the lower-order CPGs, CPG function is also influenced by sensory feedback. Thus, at least some neural circuits of the sensory-motor loop seem to be capable of producing stereotypical behaviour while sensitive to higher-order and environmental modulatory signals. Intriguingly, this state of affairs may be mapped onto the CoHSI, because a cognition or operator signal modulates a distributed and semi-autonomous system (CPG or swarm). We therefore investigate transferring neurocomputational principles to the design of HSI.

In this neuro-inspired theory that we call the \textbf{swarm-amplified human}, the swarm is seen as an extension of the human nervous system. From an operator perspective, the swarm becomes an extension of the human body \cite{Hollnagel2005,Miller2018}; an artificial body part. This shifts the focus from the design of the human-swarm channel to the interaction between human and environment which is \textit{interfaced by the swarm} \cite{LeGoc2016}. From a swarm perspective, swarm robots form behavioural subgroups ('body parts') determined by their respective local environments and human modulation while the latter is conveyed by a distributed neural pathway overlay (i.e. a flexible hop network). The research question in the context of swarm-amplified humans is: \textit{'How can the operator interact with the environment through the swarm?'}.


This perspective joins human and swarm into an adaptive, robust and scalable human-swarm loop. It highlights system adaptivity by giving the operator high-level control over his swarm 'body', as the human component is the more capable adaptive subsystem. Similar to CPGs, the swarm is capable of reacting to the environment in a closed-loop (i.e. sort out some mission goals autonomously). The swarm therefore is biased by the human component but not fully centralized to it. Robustness and scalability is preserved by the decentralized nature of the neuronal overlay.

For example, human states (e.g. cognitive states such as vigilance) may be injected into the swarm network where the human bias signal interacts with local environmental classifications of robots (e.g. heat classification). In a fire fighting scenario, a combination of $ humanClassifier_{vigilance}=high $ and $ robotSensor_{heat}=high $  could bias parts of the swarm network to flip their observable behaviour to 'danger - get out!' \cite{Penders2011,Payton2005}. This may be similar to switching from walking to running in a reactive flight response that may be an interplay of cognitive and sensory-motor computation. Note that the swarm requires no deliberate operator control and serves as a sensory organ; it is a semi-autonomous part of the human-environment loop.


\subsection{Deduced Hypotheses}

We shortly list four deduced hypotheses about how to bring together neural- and swarm principles in the interdependent human-swarm loop components. Elaboration will be given in later work due to space restrictions.

\subsubsection{Human Output - Comprehensible Passive Interaction}
It is predicted that comprehensible passive interaction is a beneficial design selection, as body coordination seems controllable and often automatic to the conscious mind. See e.g. the sense of agency \cite{Braun2018} and human state classifications \cite{Penders2011,Karavas2017}.


\subsubsection{Swarm Input - Flexible and Distributed Signal Propagation}
It is predicted that flexible and distributed signal propagation by a neural overlay (hop network) while minimizing transmitted information for robustness and scalability is a beneficial design selection. See e.g. Hebb's rule \cite{Hebb1949} that increases neural connections by a positive feedback mechanism conceptually similar to collective path selection in ants \cite{Deneubourg1990}. 


\subsubsection{Swarm Output - Inferring Global Information by Local Interactions}
It is predicted that inferred global information about the swarm and environment by local human-robot- or hub-robot interactions for preserving distributive coding is a beneficial design selection. See e.g. the virtual pheromone \cite{Payton2005} where robots summarize non-local information up the signal path conceptually similar to neural convergence \cite{Hubel1962}.  

\subsubsection{Human Input - Correlating Multi-modal Stimulation}
It is predicted that spatially and temporally correlating multi-sensory operator stimulation is a beneficial design selection, as body parts feature correlated sensory inputs. See e.g. work on tool- \cite{Miller2018} and body part integration \cite{Braun2018} as well as a multi-modal interface \cite{Penders2011}.

\section{Conclusion}

Future empirical work must explore the benefits of transferring neurocomputational principles to HSI. In any case, HSI should be seen as joining together human- and swarm capabilities in the context of system purpose.


\bibliographystyle{IEEEtran} 
\bibliography{IEEEabrv,lib}

\end{document}